\documentclass[a4paper,twoside]{article}

\usepackage{epsfig}
\usepackage{subcaption}
\usepackage{calc}
\usepackage{amssymb}
\usepackage{amstext}
\usepackage{amsmath}
\usepackage{amsthm}
\usepackage{multicol}
\usepackage{pslatex}
\usepackage{apalike}
\usepackage{SCITEPRESS}     

\usepackage[table,xcdraw]{xcolor}
\usepackage{tikz}
\usepackage[utf8]{inputenc}
\usepackage{caption}
\usepackage{hhline}

\usepackage{graphicx}
\usepackage{pgfplots}
\usepackage{cite}
\usepackage{amsfonts}
\usepackage{algorithmic}
\usepackage{textcomp}

\begin{document}

\newtheorem{definition}{Definition}
\newcommand{\encr}[1]{\left \langle {#1} \right \rangle_{\textbf{pub}}} 
\newcommand*\circled[1]{\tikz[baseline=(char.base)]{\node[shape=circle,draw,inner sep=1pt] (char) {#1};}}

\title{Security for Distributed Deep Neural Networks  \subtitle{Towards Data Confidentiality \& Intellectual Property Protection} }

\author{\authorname{Laurent Gomez\sup{1}, Marcus Wilhelm\sup{2}, Jos\'e M\'arquez\sup{3} and Patrick Duverger\sup{4}}
\affiliation{\sup{1}SAP Security Research, SAP Global Security}
\affiliation{\sup{2}Hasso Plattner Institute, University of Potsdam}
\affiliation{\sup{3}SAP Portfolio Strategy \& Technology Adoption, SAP SE}
\affiliation{\sup{4}Logistic \& IT Services, City of Antibes Juan-les-Pins}
\email{\{laurent.gomez\}@sap.com, \{marcus.wilhelm\}@student.hpi.com, \{jose.marquez\}@sap.com, \{patrick.duverger\}@antibes.com}
}


\abstract{Current developments in Enterprise Systems observe a paradigm shift, moving the needle from the backend to the edge sectors of those; by distributing data, decentralizing applications and integrating novel components seamlessly to the central systems. Distributively deployed AI capabilities will thrust this transition.\\
Several non-functional requirements arise along with these developments, security being at the center of the discussions.
Bearing those requirements in mind, hereby we propose an approach to holistically protect distributed Deep Neural Network (DNN) based/enhanced software assets, i.e. confidentiality of their input \& output data streams as well as safeguarding their Intellectual Property.\\
Making use of Fully Homomorphic Encryption (FHE), our approach enables the protection of Distributed Neural Networks, while processing encrypted data. On that respect we evaluate the feasibility of this solution on a Convolutional Neuronal Network (CNN) for image classification deployed on distributed infrastructures.
}

\onecolumn \maketitle \normalsize \vfill
\section{Introduction}
\label{sec:introduction}

\subsection{Motivation}
\label{subsec:motivation}

Until now, the backend (on-prem \& cloud) deployments were considered as the single source of truth \& unique point of access in regards of Enterprise Systems (ES). Nevertheless, a paradigm shift has been recently observed, by the deployment of ES assets towards the Edge sectors of the landscapes; by distributing data, decentralizing applications, de-abstracting technology and integrating edge components seamlessly to the central backend systems.




Capitalizing on recent advances on High Performance Computing along with the rising amounts of publicly available
labeled data, Deep Neural Networks (DNN), as an implementation of AI, have and will revolutionize virtually every current application domain as well as enable novel ones like those on 
autonomous, predictive, resilient, self-managed, adaptive, and evolving applications.




\subsection{Problem Statement}
\label{subsec:problemstatement}

Independant Software Vendors aim to protect both: data and the Intellectual Property of their AI-based software assets, deployed on potentially unsecure edge hardware \& platforms~\cite{IanGoodfellow2018}. 

The deployment of data processing capabilities throughout Distributed Enterprise Systems rises several security challenges related to the protection of input \& output data \cite{GDPR} as well as of software assets. 

In the specific context of distributed intelligence, DNN based/enhanced software will represent key investments in infrastructure, skills and governance, as well as in the acquisition of data and talents. The software industry is therefore in the direct need to safeguard these strategic investments by enforcing the protection of this new form of Intellectual Property.

\subsection{State-of-the-Art}
\label{subsection:soa}

Security of Deep Neural Networks is a current research topic taking advantage of two major cryptographic approaches: variants of Fully Homomorphic Encryption/FHE \cite{gentry2009fully} and Secure Multi-Party Computation/SMC \cite{cramer2015secure}. While FHE techniques allow addition and multiplication on encrypted data, SMC enables arithmetic operations on data shared across multi-parties.

Several approaches can be found in the literature, at different phases of the development and deployment of DNNs.

\paragraph{Secure Training}
\label{subsubse:securetraining}
Secure DNN training has been addressed using FHE \cite{graepel2012ml} and SMC \cite{shokri2015privacy}, disregarding protection once the trained model is to be productively deployed. Other Machine Learning models such as linear and logistic regressions have also been trained in a secure way in \cite{mohassel2017secureml}.
In those approaches, confidentiality of training data is guaranteed, while runtime protection (i.e. input, model, output) is out of scope.

\paragraph{Processing on Encrypted Data}
\label{subsubsec:processingonencrypteddata}
At processing phase, SMC has led to cooperative solutions where several devices work together to obtain federated inferences \cite{liu2017oblivious}, not supporting deployment of the trained DNN to trusted decentralized systems. DNN processing on FHE encrypted data is covered in CryptoNets \cite{gilad2016cryptonets}
More recently, in \cite{DBLP:journals/ijon/BoemerRL18}, the authors proposed a privacy-preserving framework for deep learning, making use of the SEAL~\cite{sealcrypto} FHE library.
While disclosure of data at runtime is prevented in these solutions, protection of DNN models remains out of the scope.

\paragraph{Intellectual Property Protection of DNN Model}
\label{subsubsec:securemodel}
In \cite{uchida2017embedding}, the authors tackles IP protection of DNN models through model watermarking. While infringement can be detected with this method, it can not be prevented. Furthermore, runtime protection of input, model and output are out of scope.

To the best of our knowledge, no other publication has holistically tackled the protection of both trained DNN models and data, targeting distributed untrusted systems.

\subsection{Data \& Intellectual Property Protection for Deep Neural Networks}
\label{subsec:proposedsolution}
In this paper we propose a novel approach for the Intellectual Property Protection of DNN-based/enhanced software while enabling data protection at processing time, making use of concepts such as Fully Homomorphic Encryption (FHE).


Once trained, DNN model parameters (i.e. weights, biases) are encrypted homomorphically. The resulting (encrypted) DNN can be distributed across untrusted landscapes, preserving its IP while mitigating the risk of reverse engineering. At runtime, FHE-encrypted insights from encrypted input data are produced by the homomorphically encrypted DNN. Confidentiality of both trained DNN, input and output data will be therefore guaranteed.

In this paper, we 
evaluate 
the overall performance (e.g. CPU, memory, disk usage) along with the accuracy of encrypted DNNs.

This paper is organized as follows: Section~\ref{sec:fundamentals} details the fundamentals of our approach. Section~\ref{sec:Approach} provides an overview of our solution. In
Sections \ref{sec:architecture} and \ref{sec:evaluation}, we present the architecture and evaluation, concluding with an outlook in Section~\ref{sec:conclusion}.

\section{Fundamentals}
\label{sec:fundamentals}

\subsection{Deep Neural Network}
\label{subsec:deepneuralnetworks}

DNNs are composed of $L$ transformation layers:
\begin{enumerate}
    \item An \textbf{input layer}, the tensor of input data $\mathbf{X}$
    \item $L-1$ \textbf{hidden layers}, mathematical computations transforming $\mathbf{X}$ sequentially.
    \item An \textbf{output layer}, the tensor of output data $\mathbf{Y}$.
\end{enumerate}  


We denote the output of layer $i$ as a tensor $\mathbf{A^{[i]}}$, with $\mathbf{A^{[0]}}=X$,and $\mathbf{A^{[L]}}=Y$. 
Tensors can have different sizes and number of dimensions. 

Each layer $\mathbf{A^{[i]}}$ depends on the mathematical computations performed at the previous layer $\mathbf{A^{[i-1]}}$.
At each layer $\mathbf{A^{[i]}}$, two types of function can be computed:
\begin{itemize}
	\item \textit{Linear}: involving polynomial operations.
	\item \textit{Non-linear}, involving non-linear operations, so called activation function, such as $max$, $exp$, $division$, ReLU, or Sigmoid. 
\end{itemize}

\subsubsection{Linear Computation Layer}
\label{subsubsec:linearoperation} 

\begin{figure}[h]%
  \centering
	\includegraphics[width=0.3\textwidth]{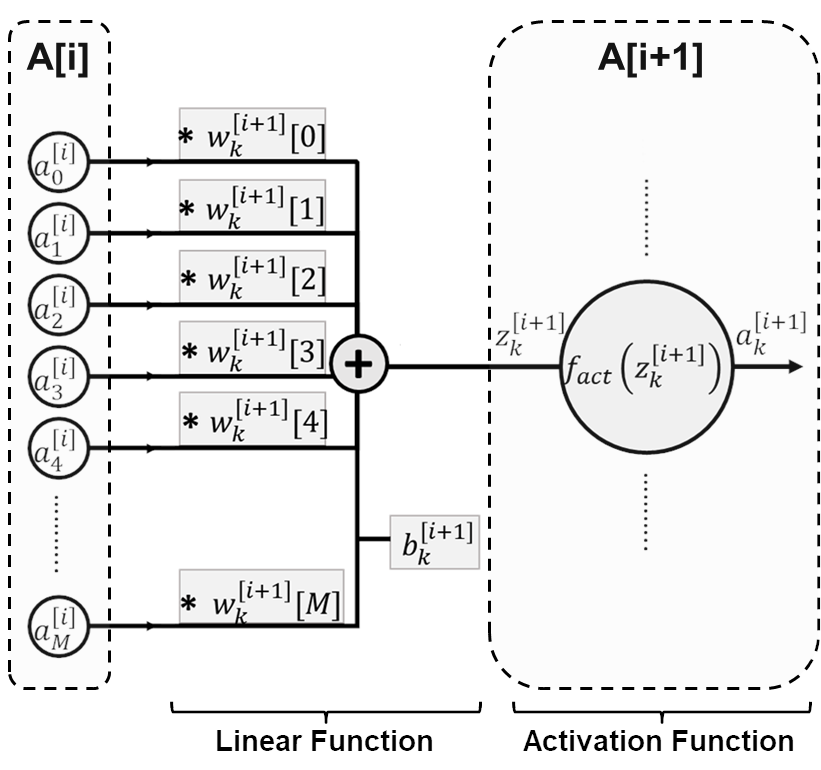}%
	\caption[Dense Layer]{Fully Connected layer with Activation Function}%
	\label{fig:denseLayer}%
\end{figure}

For the sake of clarity, we exemplify the inner linear computation with a Fully Connected (FC) layer, as depicted in Figure \ref{fig:denseLayer}.

A Fully Connected layer, noted $\mathbf{A^{[i]}}$, is composed of $n$ parallel \textit{neurons}, performing a $\mathbb{R}^n\rightarrow \mathbb{R}^n$ transformation (see Figure \ref{fig:denseLayer}). We define:

\noindent$\mathbf{a^{[i]}} =
\begin{bmatrix}
 a^{[i]}_0 \hdots  a^{[i]}_k  \hdots a^{[i]}_N
\end{bmatrix}^T$ as the output of layer $\mathrm{A}^{[i]}$;

\noindent$\mathbf{z^{[i]}} =
\begin{bmatrix}
 z^{[i]}_0 \hdots  z^{[i]}_k  \hdots z^{[i]}_N
\end{bmatrix}^T$ as the linear output of layer $\mathrm{A}^{[i]}$; 
($\mathbf{z^{[i]}}=\mathbf{a^{[i]}}$ if there is no activation function)

\noindent$\mathbf{b^{[i]}} =
 \begin{bmatrix}
 b^{[i]}_0 \hdots  b^{[i]}_k  \hdots b^{[i]}_N
 \end{bmatrix}^T$ as the bias for layer $\mathrm{A}^{[i]}$;

\noindent$\mathbf{W^{[i]}} = 
\begin{bmatrix}
 \mathbf{w^{[i]}_0}  \hdots \mathbf{w^{[i]}_k} \hdots \mathbf{w^{[i]}_N}
\end{bmatrix}^T$ as the weights for layer $\mathrm{A}^{[i]}$.

Neuron $k$ performs a linear combination of the output of the previous layer $\mathbf{a^{[i-1]}}$ multiplied by the weight vector $\mathbf{w^{[i]}_k}$ and shifted with a bias scalar $b^{[i]}_k$, obtaining the linear combination $z^{[i]}_k$:

\begin{equation}
z^{[i]}_k=\left(\sum_{l=0}^{M}w^{[i]}_k[l]*a^{[i-1]}_l\right)+b^{[i]}_k=\mathbf{w^{[i]}_k}*\mathbf{a^{[i-1]}}+b^{[i]}_k
\end{equation}

Vectorizing the operations for all the neurons in layer $A^{[i]}$ we obtain the dense layer transformation:
\begin{equation}
\mathbf{z^{[i]}}=\mathbf{W^{[i]}}*\mathbf{a^{[i-1]}}+\mathbf{b^{[i]}}\\
\end{equation}
where $\mathbf{W}$ and $\mathbf{b}$ are the parameters for layer $A^{[i]}$.

\subsubsection{Activation Functions}

Activation functions are the major source of non-linearity in DNNs. They are performed element-wise ($\mathbb{R}^0\rightarrow \mathbb{R}^0$, thus easily vectorized), and are generally located after linear transformations such as Fully Connected layers.

\begin{equation}
a^{[i]}_k=f_{act}\left(z^{[i]}_k\right)
\end{equation}

Several activation functions have been proposed in the literature but \textit{Rectified Linear Unit (ReLU)} is currently considered as the most efficient activation function for DL. 
Several variants of ReLU exist, such as Leaky ReLU\cite{maas2013rectifier}, ELU\cite{clevert2015fast} or its differentiable version \textit{Softplus}.
\begin{equation}
\begin{split}
  ReLU(z)&=z^+=max(0, z) \\
  Softplus(z)&= log(e^z + 1)
  \end{split}
\end{equation}



\subsection{Homomorphic Encryption} \label{subsec:s31_homomorphicEnc}

While preserving data privacy, Homomorphic Encryption (HE) schemes allow certain computations on ciphertext without revealing neither its inputs nor its internal states.
Gentry \cite{gentry2009fully} first proposed a Fully Homomorphic Encryption (FHE) scheme, which theoretically could compute any kind of arithmetic circuit, but is computationally intractable in practice. FHE evolved into more efficient schemes preserving addition and multiplication over encrypted data, such as BGV \cite{cryptoeprint:2011:277}, FV \cite{FV} or CKKS \cite{DBLP:journals/iacr/CheonHKKS18}, allowing approximations of multiplicative inverse, exponential and logistic function, or discrete Fourier transformation. Similar to asymmetric encryption, a public-private key pair (\textit{pub}, 
\textit{priv}) is generated.

\begin{definition}
	An encryption scheme is called homomorphic over an operation $\odot$ if it supports the following 
	\begin{gather*}
		Enc_{\mathbf{pub}}(m) = \encr{m}, \forall m \in \mathcal{M} \\
		\encr{m_1\odot m_2} = \encr{m_1} \odot \encr{m_2}, \forall m_1, m_2 \in \mathcal{M}
	\end{gather*}
	where $Enc_{\mathbf{pub}}$ is the encryption algorithm and $\mathcal{M}$ is the set of all possible messages.
\end{definition}

\begin{definition}
	Decryption is performed as follows
	\begin{gather*}
		Enc_{\mathbf{pub}}(m) = \encr{m}, \forall m \in \mathcal{M} \\
		Dec_{\mathbf{priv}}(\encr{m}) = m
	\end{gather*}
	where $Dec_{\mathbf{priv}}$ is the decryption algorithm and $\mathcal{M}$ is the set of all possible messages.
\end{definition}


\subsection{Challenges}
\label{subsec:challenges}

Even though HE schemes seem theoretically promising, their usage comes with several drawbacks, particularly when applied to Deep Learning.

\subsubsection{Noise budget}
\label{subsubsec:noisebudget}
In Gentry's lattice-based HE schemes\cite{gentry2009fully} and subsequent variants of it, ciphertexts contain a small term of random noise drawn from some probability distribution.
To estimate the current magnitude of noise, a \textbf{noise budget} can be calculated, that starts as a positive integer, decreases with subsequent operations and reaches 0 exactly when the ciphertext becomes indecipherable.
The noise budget is more strongly affected by multiplications as by additions.

In order to cope with that challenge, encryption parameters can be adjusted accordingly to the required computation depth of an arithmetic circuit.

\subsubsection{FHE libraries and APIs}
\label{subsubsec:heterogeonityinFHElibrariesandAPIs}

Multiple FHE libraries are available \cite{halevi2014algorithms}, \cite{PALISADE}, \cite{sealcrypto}, \cite{ducas2015fhew}.
Depending on the supported HE schemes, those libraries show noticeable difference on performance (e.g. computational, memory consumption), on supported operations type (e.g. addition, multiplication, negative, square, division), datatype (e.g. floating point, integer), and chipset infrastructure (e.g. CPU, GPU).

In addition, and regardless on their level of maturity and performance, HE libraries can be configured through several encryption parameters such as: 
\begin{itemize}
	\item{Polynomial degree or modulus}: which determines the available noise budget and strongly affects the performance.
    \item{Plaintext modulus}: which is mostly associated to the size of input data.
	\item{Security parameter}: which sets the reached level of security in bits of the cryptosystem (e.g. 128, 192, 256-bit security level).
\end{itemize}

Fine-tuning of those encryption parameters enables developers to optimize the performance of encryption and encrypted operations. The selection of the right encryption parameters depends on the size of the plaintext data, targeted accuracy loss or level of security.

\subsubsection{Linear function support only}
\label{subsubsec:nonlinearfunctionapproximation}
By construction, linear functions, composed of addition and multiplication operations, are seamlessly protected by FHE.
But, non-linear activation functions such as ReLU or Sigmoid require approximation to be computed with FHE schemes.

The challenge lies on the transformation of activation functions into polynomial approximations
supported by HE schemes. We elaborate more on approximation of activation functions in Section \ref{subsec:activationlayerprotection}.

\subsubsection{Supported plaintext type}
\label{subsubsec:supportedplaintexttype}
The vast majority of HE schemes allow operations on integers \cite{halevi2014algorithms,sealcrypto}
, while others use booleans \cite{cryptoeprint:2018:421} or floating point numbers \cite{DBLP:journals/iacr/CheonHKKS18,sealcrypto}. 
In the case of integer supporting HE schemes, rational numbers can be approximated using fixed-point arithmetic by scaling with a scaling factor and rounding.

\subsubsection{Performance}
\label{subsubsec:performance}

FHE schemes are computationally expensive and memory consuming. 
In addition, ciphertexts are often significantly bigger than plaintexts and thus use more memory and disk space.

Even if in the past years the performance of FHE made it impractical, recent FHE schemes show promising throughput. New FHE libraries take also advantage of GPU acceleration.


In addition, modern implementations of HE schemes such as \textbf{HELib}~\cite{halevi2014algorithms}, \textbf{SEAL}~\cite{sealcrypto}, or \textbf{PALISADE}~\cite{PALISADE} benefit from Single Instruction Multiple Data (SIMD), allowing multiple integers to be stored in a single ciphertext and vectorizing operations, which can accelerate certain applications significantly.

\section{Approach}
\label{sec:Approach}

As introduced in Section~\ref{subsec:problemstatement}, the delivery of DNN-enriched insights come at a cost. ISVs aim to guarantee data security, together with the IP protection of their DNN-based software assets, deployed on potentially unsecure edge hardware \& platforms. In order to achieve those security objectives on DNN, we utilize FHE schemes to operate on ciphertext at runtime. 

Consequently, secure training of DNN is out of scope of our approach as we focus on runtime execution. We assume that DNN training already preserves both data privacy and confidentiality, and the resulting trained model. Once a model is trained, as discussed in Section~\ref{subsubsec:linearoperation}, we obtain a set of parameters for each DNN layer; i.e weights $\mathbf{W^{[i]}}$ and biases $\mathbf{b^{[i]}}$.
Those parameters constitute the IP to be protected when deploying a DNN to distributed systems.



\subsection{Linear Computation Layer Protection} 
\label{subsec:deepneuralnetworkencryption}

Our approach is agnostic from the type of layer. In \cite{DBLP:conf/icete/GomezIMD18}, we detail the encryption of layers such as Convolutional Layer or Batch Normalization.
For sake of simplicity, we exemplify the encryption of DNN layers parameters on FC layers. 
Since FC are simply a linear transformation on the previous layer's outputs, encryption is achieved straightforwardly as follows
\begin{equation}
\begin{split}
  \encr{\mathbf{z^{[i]}}}& = \encr{\mathbf{W^{[i]}}*\mathbf{a^{[i-1]}}+\mathbf{b^{[i]}}} \\
												 & = \encr{\mathbf{W^{[i]}}}*\encr{\mathbf{a^{[i-1]}}}+\encr{\mathbf{b^{[i]}}} \\
\end{split}
\end{equation}

\subsection{Activation Function Protection}
\label{subsec:activationlayerprotection}
Due to their innate non-linearity, activation functions need to be approximated with polynomials to be encrypted with FHE.
Several approaches have been elaborated in the literature. In \cite{NIPS2014_5267} and \cite{gilad2016cryptonets}, the authors proposed to use a square function as activation function. The last layer, a sigmoid activation function, is only applied during training.
Chabanne et al. used Taylor polynomials around $x=0$, studying performance based on the polynomial degree~\cite{chabanne2017privacy}. In \cite{hesamifard2017cryptodl}, Hesamifard et al. approximate instead the derivative of the function and then integrate to obtain their approximation.

Regardless on the approximation technique, we denote activation function $f_{act}()$ approximation as 
\begin{equation}
\mathbf{f_{act}()} \approx \mathbf{f_{approxact}()}
\end{equation}

By construction, we have  
\begin{equation}
\begin{split}
\encr{\mathbf{a^{[i]}_k}}& = \encr{\mathbf{f_{act}\left(z^{[i]}_k\right)}}\\
												 & \equiv \encr{\mathbf{f_{approxact}\left(z^{[i]}_k\right)}}
\end{split}
\end{equation}

\subsection{Architecture}
\label{sec:architecture}

\begin{figure}[ht]%
  \centering
	\includegraphics[width=0.35\textwidth]{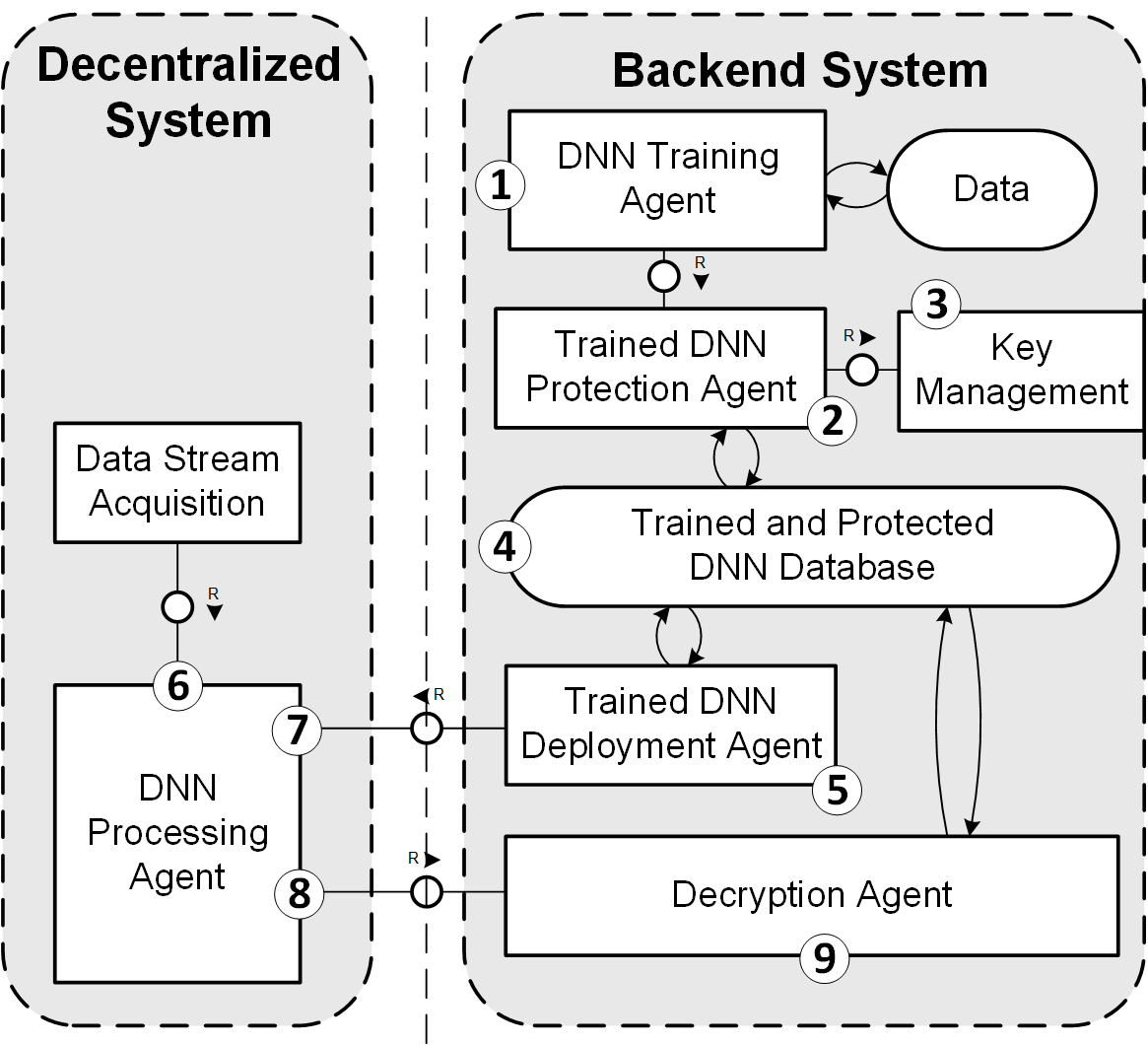}%
	\caption[ProcDiag]{Overall Architecture}%
	\label{fig:prDiag}%
\end{figure}

In this section we outline the architecture of our IP protection system, as depicted in Figure \ref{fig:prDiag}.

\subsubsection{Encryption of trained DNN}
\label{subsection:encryptionoftraineddnn}
In the backend, a DNN is trained within a \textit{DNN Training Agent}, \circled{1}. The outcome of the training (NN architecture and parameters) is pushed to the \textit{Trained DNN Protection Agent}, \circled{2}. Alternatively, an already trained DNN can be imported directly into the \textit{Protection Agent}.
The \textit{DNN Protection Agent} generates a Fully Homomorphic key pair from the \textit{Key Generator} component, \circled{3}. The DNN is then encrypted and stored together with its homomorphic key pair in the \textit{Trained and Protected DNN Database}, \circled{4}.

\subsubsection{Deployment of trained and protected DNN}
\label{subsection:deploymentoftrainedandprotecteddnn}
At the deployment phase, the \textit{Trained DNN Deployment Agent} deploys the DNN on distributed systems, together with its public key, \circled{5}.

\subsubsection{DNN processing}
\label{subsection:dnnprocessing}        
On the distributed system, data is collected by a \textit{Data Stream Acquisition} component, \circled{6}, and forwarded to the \textit{DNN Processing Agent}, \circled{7}. The input layer does not involve any computation, and therefore can be seamlessly FHE encrypted as follows:

\begin{equation}
\mathbf{X} \xrightarrow{encryption} Enc_{\mathbf{pub}} (X) = \encr{X}
\end{equation}

Encrypted inferences are sent to the \textit{Decryption Agent}, \circled{8}, for their decryption using the private key associated to the DNN, \circled{9}. FHE encryption propagates across the DNN layers, from the input to the output layer. By construction, the output layer is therefore encrypted homomorphically.

The decryption of the last layer's output $\mathbf{Y}$ is done with the private key \textit{priv}:

\begin{equation} \encr{\mathbf{A^{[L]}}} \xrightarrow{decryption} Dec_{\mathbf{priv}}\left(\encr{\mathbf{A^{[L]}}}\right) =\mathbf{Y}
\end{equation}

The Intellectual Property of the DNN, together with the input \& output results, is protected from any disclosure on the distributed system throughout the entire process.

\section{Evaluation}
\label{sec:evaluation}

As detailed in Section~\ref{subsec:challenges}, FHE introduces additional computational costs at each step of the DNN life-cycle.
In this section, we evaluate performance overhead from computation time, memory load and disk usage perspectives at DNN model and processing encryption and output decryption.

\subsection{Hardware Setup}
\label{subsection:hardwaresetup}

%

As \textit{backend}, we use a NVIDIA DGX-1\footnote{https://www.nvidia.com/en-us/data-center/dgx-1/} server, empowered with 8 Tesla V100 GPUs. This machine is theoretically not resource-constrained (computation \& memory). We reasonably neglect the impact of the performance overhead introduced by FHE on DNN trained model encryption and output decryption. 

We deploy and execute our encrypted DNN on a NVIDIA Jetson-TX2\footnote{https://www.nvidia.com/en-us/autonomous-machines/embedded-systems-dev-kits-modules/}. Powered by NVIDIA Pascal architecture, this platform embeds 256 CUDA cores, CPU HMP Dual Denver 2\/2 MB L2 + Quad ARM® A57\/2 MB L2, and 8 GB of memory. This platform gets closer to the hardware configuration of a Distributed Enterprise System.

\subsection{Software Setup} 
\label{subsection:softwaresetup}

\paragraph{DNN Model}

As demonstrated in Section~\ref{sec:Approach}, our approach is fully agnostic from NN topology, or implementation.
For the sake of our evaluation, involving several modifications to the NN model, we choose a simple CNN classifier\footnote{https://github.com/keras-team/keras/blob/master/examples/cifar10\_cnn.py}, implemented with the Keras library\footnote{https://keras.io}. Two datasets have been used in our experiment: CIFAR10\footnote{https://www.cs.toronto.edu/~kriz/cifar.html}, for image classification, and MNIST\footnote{http://yann.lecun.com/exdb/mnist/} for handwritten digits classification.

\begin{figure}[ht]
  \centering
	\includegraphics[width=0.35\textwidth]{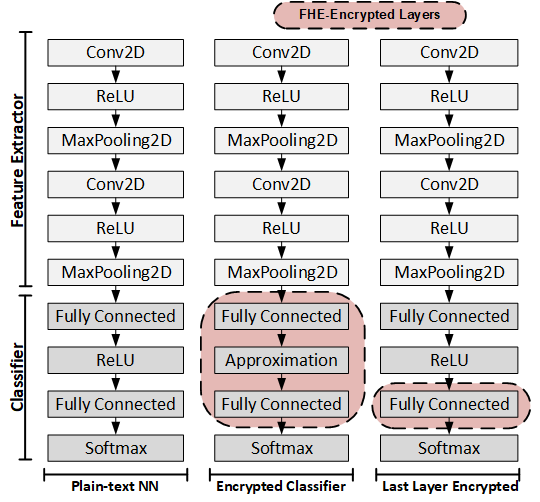}
	\caption[KerasCNN]{Keras Convolutional Neural Network.}
	\label{fig:KerasCNN}
\end{figure}

As depicted in Figure \ref{fig:KerasCNN}, we distinguish two main parts in this CNN: a \textit{feature extractor} and a \textit{classifier}.
The feature extractor reduces the amount of information from the input image, into a set of high level and more manageable features. This step facilitates the subsequent classification of the input data. 

Composed of four layers, $[\textit{FC} \rightarrow \textit{ReLU} \rightarrow \textit{FC} \rightarrow \textit{Softmax}]$, the classifier categorizes the input data according to the extracted features, and outputs discrete probability distribution over 10 classes of objects. 

As reference point, we evaluate key performance figures at model training and processing time without encryption.
Once trained, the size of the CNN plaintext model is 9.6Mb. 
On Jetson TX2, single unencrypted image classification is computed on average in 89.1ms.

Performances vary from 2.1s for an encrypted classification, with only 53.9Mb consumed memory, up to 1h33m with almost 5Gb of consumed memory.


%
%



\paragraph{FHE library}

As introduced in section \ref{subsec:challenges}, several libraries are available for FHE.
We use SEAL 
\cite{sealcrypto} C library from Microsoft Research running on CPU.
This choice is motivated by the library's performance, support of multiple schemes such as BGV \cite{cryptoeprint:2011:277}, stability, and documentation.
The use of SEAL, implemented in C++, with the Keras Python library requires some engineering efforts.
To enable both fast performance of the native C++ library and rapid prototyping using Python, we use Cython\footnote{https://cython.org/}.


We conduct our evaluation with the BGV scheme \cite{cryptoeprint:2011:277}, utilizing the integer encoding with SIMD support. To handle the floating-point DNN parameters, we use fixed-point arithmetic with a fixed scaling factor, similarly to CryptoNets\cite{gilad2016cryptonets}. This has no noticeable impact on the classification accuracy, if a suitable scaling factor is applied. 
The SIMD operations allow for optimized performance through vectorization. 




\subsection{Linearization}
\label{subsection:linearization}

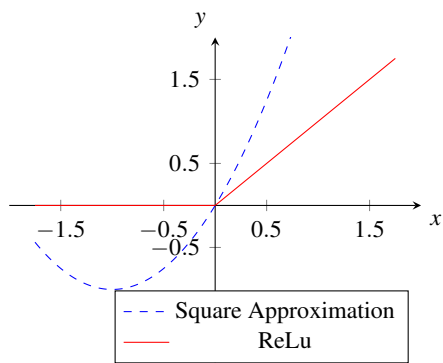
\begin{figure}
\centering
\begin{tikzpicture}[>=latex]
\begin{axis}[
  axis x line=center,
  axis y line=center,
	font=\footnotesize,
  xtick={-1.5,...,1.5},
  ytick={-1.5,...,1.5},
  xlabel={$x$},
  ylabel={$y$},
  xlabel style={below right},
  ylabel style={above left},
  xmin=-2,
  xmax=2,
  ymin=-2,
  ymax=2,
	width=7cm,
	legend pos=south east]
		\addplot[dashed, mark=none,color=blue,domain=-1.75:1.75, samples=500] {x^2+2*x};
		\addplot [mark=none,color=red,domain=-1.75:0] {0};
		\addplot [mark=none,color=red,domain=0:1.75] {x};
		\legend{Square Approximation, ReLu}

\end{axis}
\end{tikzpicture}
\caption{ReLU Approximation as Square Function}
\label{fig:squareapproximation}
\end{figure}

We tackle the problem of linearization of the ReLU functions following approaches: we approximate it with a modified square function, and we skip activation function.
The modified square function $x^2+2x$ (see Figure~\ref{fig:squareapproximation}) is derived from the ReLU approximation proposed in \cite{chabanne2017privacy}. In order to optimize the computation of that function on ciphertexts, we used simpler coefficients.

\begin{figure}[ht]
\centering
\begin{tikzpicture}
\begin{axis}[
    xlabel={Epoch},
    ylabel={Validation accuracy},
    xmin=0, xmax=200,
    ymin=0.96, ymax=0.995,
    xtick={0,20,40,60,80,100,120,140,160,180,199},
    legend pos=south east,
    ymajorgrids=true,
    grid style=dashed,
    width=7cm
]

\addplot[
    color=green,
		dashed
    ]
    table [x=step, y=value, col sep=comma] {images/csv/mnist/original/val_acc.csv};
\addplot[
    color=blue,
		dotted
    ]
    table [x=step, y=value, col sep=comma] {images/csv/mnist/1noacti/val_acc.csv};
    \legend{Original model,No activ. in last layer}
\addplot[
    color=red,
    ]
    table [x=step, y=value, col sep=comma] {images/csv/mnist/1xsquareplustwox/val_acc.csv};
    \legend{Original model,No activation function,$x^2+2x$}

    \end{axis}
\end{tikzpicture}
\caption{Classification Accuracy with ReLU Approximation - MNIST Dataset.}
\label{fig:ActivationFunctionApproximationMNIST}
\end{figure}
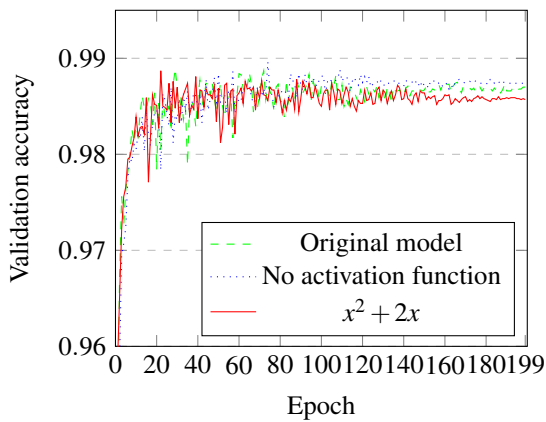

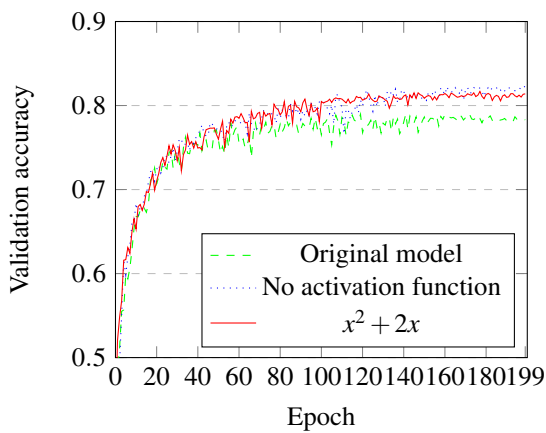
\begin{figure}[ht]
\centering

\begin{tikzpicture}
\begin{axis}[
    xlabel={Epoch},
    ylabel={Validation accuracy},
    xmin=0, xmax=200,
    ymin=0.5, ymax=0.9,
    xtick={0,20,40,60,80,100,120,140,160,180,199},
    legend pos=south east,
    ymajorgrids=true,
    grid style=dashed,
    width=7cm
]

\addplot[
    color=green,
		dashed
    ]
    table [x=step, y=value, col sep=comma] {images/csv/cifar/original/val_acc.csv};
\addplot[
    color=blue,
		dotted
    ]
    table [x=step, y=value, col sep=comma] {images/csv/cifar/1noacti/val_acc.csv};
    \legend{Original model,No activ. in last layer}
\addplot[
    color=red,
    ]
    table [x=step, y=value, col sep=comma] {images/csv/cifar/1xsquareplustwox/val_acc.csv};
    \legend{Original model,No activation function,$x^2+2x$}

    \end{axis}
\end{tikzpicture}
\caption{Classification Accuracy with ReLU Approximation - CIFAR10 Dataset.}
\label{fig:ActivationFunctionApproximationCIFAR10}
\end{figure}

In order evaluate the impact of these approaches, we trained the CNN on the CIFAR10 and MNIST datasets, replacing the last ReLU activation. 
Depicted in Figure~\ref{fig:ActivationFunctionApproximationMNIST} and Figure~\ref{fig:ActivationFunctionApproximationCIFAR10}, we report the accuracy loss. Both approximations have merely a minor impact on the output classification accuracy.

Skipping the last activation function shows good results on this simple CNN, but we do not want to generalize to any other DNN or dataset.

\subsection{Experimentation Results}
\label{subsection:experimentationresults}

\subsubsection{Model \& Data Protection}

Intellectual Property-wise, we consider the feature extractor as of minor importance, as CNNs generally use state of the art feature extractor. 
The IP of the model rather lies in the parameters, weights and bias, of the trained classifier.
For that reason, we encrypt the classifier only, as a first step towards full model encryption, as depicted in Figure ~\ref{fig:KerasCNN}.
To better understand the impact of computation depth, we also complete our evaluation with the encryption of the last FC layer only.

Confidentiality-wise, we evaluate the impact of extracted features encryption by comparing processing performance on an encrypted model with plaintext and encrypted feature extractor outputs.

As depicted in Figure \ref{fig:KerasCNN}, we evaluate our approach on three modified versions of the model:
\begin{itemize}
\item{Last FC Layer Encrypted} 
\item{Full Classifier Encrypted with no Activation Function}
\item{Full Classified Encrypted with our Modified Square Activation Function}
\end{itemize}

Confidentiality-wise, we evaluate the impact of extracted features encryption by comparing processing performance on an encrypted model with plaintext and encrypted feature extractor outputs.

In order to optimize our approach, we omit the \textit{Softmax} layer within the classifier. This layer does not have any influence on the classification results, as \textit{Softmax} layer is mostly required at training phase, to normalize network outputs probability distribution, for more consistent loss calculations.

The overall experiment as described in section~\ref{sec:architecture} has been applied 5 times on each model.
We report average evaluation metrics for each step: model encryption, processing encryption and decryption.



\subsubsection{DNN Model Encryption}
\label{subsubsection:modelencryption}

Each trained CNN model is encrypted on DGX-1's CPU.
In Table~\ref{tab:modelencryption}, we depict the resource consumption average on the following metrics:
\begin{itemize}
\item{Time to Compute}: Time to encrypt the model. 
\item{Model Size}: Size of resulting encrypted model.
\item{Memory Load}: Overall memory usage for model encryption.
\end{itemize}

We target three security levels: 128, 192, and 256-bits.
For each of those, we optimize SEAL parameters as introduced in section \ref{subsec:challenges}, maximizing performance, and minimizing leftover noise budget.

Compared to the plaintext model size (9.6Mb), encrypted model size increases by a factor of 8,22 in the best case.

\subsubsection{DNN Processing Encryption}
\label{subsubsection:DNNprocessingencryption}

The three encrypted CNN models deployed on Jetson-TX2 for CPU based encrypted processing.
At this stage, we evaluate the following metrics 
\begin{itemize}
\item{Time to compute}: Processing time for an encrypted classification.
\item{Memory}: Memory usage for encrypted classification. 
\item{Remaining Noise Budget}: At the end of processing encryption, we evaluate the remaining noise budget, which determines if additional encryption operations could be performed on the output vector.
\end{itemize}

In Table \ref{tab:processingencryption} and Table \ref{tab:processingencryptionwithencryptedinput}, we depict the performance of encrypted processing with plaintext and encrypted previous layer outputs.
We study the impact of confidentality preservation of the preceding layer outputs.
SEAL library supports secure computation over plaintext and ciphertext producing ciphertext. As a consequence, output of the last MaxPooling2D layer can be processed in FHE-encrypted Fully Connected layer. Secure computation between plaintext and ciphertext has a lower impact on performance.

We observe a slight performance improvement on time to compute and memory between 128 and 192-bit security level. This is due to the FHE parameters optimization as described in Section~\ref{subsubsection:modelencryption}, where initial noise budget is oversized for 128-bit security level, which has a direct impact to performance.

Experiment results show that, depending on the level of achieved security, and targeted scenario, we can achieve at best encrypted classification in 2.1s (for 128 level security and only one layer encrypted). 


\begin{table*}[!ht]
\centering
\begin{tabular}{l|l|l|l|l|l|l|l|l|l|}
\cline{2-10}
                                               & \multicolumn{9}{c|}{\textbf{Achieved Security Level (bits)}}                                                                                                                                                                                                                                                                                                                                                                                                                      \\ \hhline{~|---------|}
                                               & \multicolumn{1}{c|}{\cellcolor[HTML]{EFEFEF}\textit{128}} & \multicolumn{1}{c|}{\cellcolor[HTML]{EFEFEF}\textit{192}} & \multicolumn{1}{c|}{\cellcolor[HTML]{EFEFEF}\textit{256}} & \multicolumn{1}{c|}{\textit{128}} & \multicolumn{1}{c|}{\textit{192}} & \multicolumn{1}{c|}{\textit{256}} & \multicolumn{1}{c|}{\cellcolor[HTML]{EFEFEF}\textit{128}} & \multicolumn{1}{c|}{\cellcolor[HTML]{EFEFEF}\textit{192}} & \multicolumn{1}{c|}{\cellcolor[HTML]{EFEFEF}\textit{256}} \\ \hhline{~|---------|}
                                               & \multicolumn{3}{c|}{\cellcolor[HTML]{EFEFEF}\textit{Full Classifier - $x^2+2x$}}                                                                                                & \multicolumn{3}{c|}{\textit{Full Classifier - No Act.}}                                                    & \multicolumn{3}{c|}{\cellcolor[HTML]{EFEFEF}\textit{Last Layer}}                                                                                                                  \\ \hline
\multicolumn{1}{|l|}{\textbf{Time to Compute (s)}} & \cellcolor[HTML]{EFEFEF}287.2                            & \cellcolor[HTML]{EFEFEF}174.6                            & \cellcolor[HTML]{EFEFEF}1221.3                           & 43.7                             & 32.9                             & 90.2                             & \cellcolor[HTML]{EFEFEF}2.1                              & \cellcolor[HTML]{EFEFEF}2.1                            & \cellcolor[HTML]{EFEFEF}4.5                              \\ \hline
\multicolumn{1}{|l|}{\textbf{Memory Load (Mb)}}          & \cellcolor[HTML]{EFEFEF}4683.1                          & \cellcolor[HTML]{EFEFEF}2924.3                          & \cellcolor[HTML]{EFEFEF}4899.2                          & 1162.5                          & 869.2                           & 2342.3                          & \cellcolor[HTML]{EFEFEF}53.9                            & \cellcolor[HTML]{EFEFEF}54.0                            & \cellcolor[HTML]{EFEFEF}117.7                           \\ \hline
\multicolumn{1}{|l|}{\textbf{Remaining Noise Budget}} & \cellcolor[HTML]{EFEFEF}221.6                             & \cellcolor[HTML]{EFEFEF}80.2                              & \cellcolor[HTML]{EFEFEF}88.8                              & 91.4                              & 23.8                              & 16.2                              & \cellcolor[HTML]{EFEFEF}58.8                              & \cellcolor[HTML]{EFEFEF}20.2                              & \cellcolor[HTML]{EFEFEF}60.2                              \\ \hline
\end{tabular}
\caption{Runtime Encryption with Plaintext Input}
\label{tab:processingencryption}
\end{table*}

\begin{table*}[!ht]
\centering
\begin{tabular}{l|l|l|l|l|l|l|l|l|l|}
\cline{2-10}
                                               & \multicolumn{9}{c|}{\textbf{Achieved Security Level (bits)}}                                                                                                                                                                                                                                                                                                                                                                                                                      \\ \hhline{~|---------|}
                                               & \multicolumn{1}{c|}{\cellcolor[HTML]{EFEFEF}\textit{128}} & \multicolumn{1}{c|}{\cellcolor[HTML]{EFEFEF}\textit{192}} & \multicolumn{1}{c|}{\cellcolor[HTML]{EFEFEF}\textit{256}} & \multicolumn{1}{c|}{\textit{128}} & \multicolumn{1}{c|}{\textit{192}} & \multicolumn{1}{c|}{\textit{256}} & \multicolumn{1}{c|}{\cellcolor[HTML]{EFEFEF}\textit{128}} & \multicolumn{1}{c|}{\cellcolor[HTML]{EFEFEF}\textit{192}} & \multicolumn{1}{c|}{\cellcolor[HTML]{EFEFEF}\textit{256}} \\ \hhline{~|---------|}
                                               & \multicolumn{3}{c|}{\cellcolor[HTML]{EFEFEF}\textit{Full Classifier - $x^2+2x$}}                                                                                                & \multicolumn{3}{c|}{\textit{Full Classifier - No Act.}}                                                    & \multicolumn{3}{c|}{\cellcolor[HTML]{EFEFEF}\textit{Last Layer}}                                                                                                                  \\ \hline
\multicolumn{1}{|l|}{\textbf{Time to Compute (s)}} & \cellcolor[HTML]{EFEFEF}2902.6                           & \cellcolor[HTML]{EFEFEF}1048.6                           & \cellcolor[HTML]{EFEFEF}5627                           & 367.1                            & 272.2                            & 835.9                            & \cellcolor[HTML]{EFEFEF}19.2                             & \cellcolor[HTML]{EFEFEF}19.2                             & \cellcolor[HTML]{EFEFEF}40.2                             \\ \hline
\multicolumn{1}{|l|}{\textbf{Memory Load (Mb)}}          & \cellcolor[HTML]{EFEFEF}5047.8                          & \cellcolor[HTML]{EFEFEF}5809                          & \cellcolor[HTML]{EFEFEF}4906                          & 2314.8                          & 1733.7                          & 4644.5                          & \cellcolor[HTML]{EFEFEF}119.5                           & \cellcolor[HTML]{EFEFEF}118.5                           & \cellcolor[HTML]{EFEFEF}253                           \\ \hline
\multicolumn{1}{|l|}{\textbf{Remaining Noise Budget}} & \cellcolor[HTML]{EFEFEF}206.0                             & \cellcolor[HTML]{EFEFEF}65.0                              & \cellcolor[HTML]{EFEFEF}76.0                              & 79.0                              & 11.0                              & 3.0                               & \cellcolor[HTML]{EFEFEF}45.0                              & \cellcolor[HTML]{EFEFEF}10.0                              & \cellcolor[HTML]{EFEFEF}52.0                              \\ \hline
\end{tabular}
\caption{Runtime Encryption with Encrypted Input}
\label{tab:processingencryptionwithencryptedinput}
\end{table*}

\subsubsection{Decryption}
\label{subsubsection:decryption}

Following our approach, encrypted output are decrypted by the backend, on DGX-1. We therefore consider decryption as not computationally expensive, compared to encryption.
Results are available in Table \ref{tab:decryptionprocessing}.

\begin{table*}[!ht]
\centering
\begin{tabular}{l|c|l|l|l|l|l|l|l|l|}
\cline{2-10}
                                               & \multicolumn{9}{c|}{\textbf{Achieved Security Level (bits)}}                                                                                                                                                                                                                                                                                                                                                                                                                 \\ \hhline{~|---------|}
                                               & \cellcolor[HTML]{EFEFEF}\textit{128}                 & \multicolumn{1}{c|}{\cellcolor[HTML]{EFEFEF}\textit{192}} & \multicolumn{1}{c|}{\cellcolor[HTML]{EFEFEF}\textit{256}} & \multicolumn{1}{c|}{\textit{128}} & \multicolumn{1}{c|}{\textit{192}} & \multicolumn{1}{c|}{\textit{256}} & \multicolumn{1}{c|}{\cellcolor[HTML]{EFEFEF}\textit{128}} & \multicolumn{1}{c|}{\cellcolor[HTML]{EFEFEF}\textit{192}} & \multicolumn{1}{c|}{\cellcolor[HTML]{EFEFEF}\textit{256}} \\ \hhline{~|---------|}
                                               & \multicolumn{3}{c|}{\cellcolor[HTML]{EFEFEF}\textit{Full Classifier - $x^2+2x$}}                                                                                           & \multicolumn{3}{c|}{\textit{Full Classifier - No Act.}}                                                    & \multicolumn{3}{c|}{\cellcolor[HTML]{EFEFEF}\textit{Last Layer}}                                                                                                                  \\ \hline
\multicolumn{1}{|l|}{\textbf{Time to Compute (s)}} & \multicolumn{1}{l|}{\cellcolor[HTML]{EFEFEF}2.9}    & \cellcolor[HTML]{EFEFEF}1.7                              & \cellcolor[HTML]{EFEFEF}3.2                              & 0.6                              & 0.6                              & 1.0                              & \cellcolor[HTML]{EFEFEF}0.2                              & \cellcolor[HTML]{EFEFEF}0.1                              & \cellcolor[HTML]{EFEFEF}0.2                              \\ \hline
\multicolumn{1}{|l|}{\textbf{Memory Load (Mb)}}          & \multicolumn{1}{l|}{\cellcolor[HTML]{EFEFEF}963.8} & \cellcolor[HTML]{EFEFEF}397.4                           & \cellcolor[HTML]{EFEFEF}2062.5                          & 123.4                           & 73.4                           & 267.1                          & \cellcolor[HTML]{EFEFEF}17.8                           & \cellcolor[HTML]{EFEFEF}17.8                            & \cellcolor[HTML]{EFEFEF}38.7                           \\ \hline
\end{tabular}
\caption{Decryption - Performance}
\label{tab:decryptionprocessing}
\end{table*}

\section{Conclusion}
\label{sec:conclusion}

In this paper, we discuss and evaluate a holistic approach for the protection of distributed Deep Neural Network (DNN) enhanced software assets, i.e. confidentiality of their input \& output data streams as well as safeguarding their Intellectual Property. On that matter, we take advantage of Fully Homomorphic Encryption (FHE). We evaluate the feasibility of this solution on a Convolutional Neural Network (CNN) for image classification.

Our evaluation on NVIDIA DGX-1 and Jetson-TX2 shows promising results on the CNN image classifier. 
Performances vary from 2.1s for an encrypted classification, with only 53.9Mb consumed memory, up to 1h33m with almost 5Gb of consumed memory.


\bibliographystyle{apalike}
{\small
\bibliography{./bib/Bibliography}}

\end{document}